\documentclass{ltxguide}

\usepackage{color,graphicx,shortvrb}
\usepackage{amsmath,amsthm}
\usepackage{amscd}

\makeindex % generate index data

\chardef\bslash=`\\ % p. 424, TeXbook
\theoremstyle{definition}

\theoremstyle{remark}

\newcommand{\eval}[2][\right]{\relax
  \ifx#1\right\relax \left.\fi#2#1\rvert}

\newcommand{\envert}[1]{\left\lvert#1\right\rvert}

\DeclareGraphicsExtensions{.ps}
\let\package\textsf

\newlength{\gxlen}
\MakeShortVerb{\|}

\begin{document}

\title{ Percolation  of New Product Critical  Market Penetration}

\author{   Armando Barra\~n\'on  
\footnote{ Universidad Aut\'onoma Metropolitana. Unidad Azcapotzalco.
Av. San Pablo 124, Col. Reynosa-Tamaulipas, Mexico City. email: bca@correo.azc.uam.mx } ;
 } 

\def\rightmark{Percolation  of New Product Critical  Market Penetration}
\def\leftmark{A. Barra\~n\'on et al.}

\date{10 October 2003}

\maketitle

\abstract

  A simulation of new product market penetration in a social 
environment is performed, using a spintronic model, where
 each element of a 3D network interacts with its first neighbors.  
Agents are assumed to be rational, with a perfect market foresight.
 Unitary production cost  decreases when consumption is increased. 
Simulations indicate that social interaction is the most important factor
 for new product market penetration as compared to the consumer
 readiness to pay a higher price. Besides the critical exponent of the
 nucleation of new consumers  is computed, signing a phase transition
 characterized by the build up of new consumers intermediate clusters
 and this critical exponent is compared to others belonging to several
 critical phenomena. 

\section{ Introduction}
   The importance of building up groups of consumers of brand new 
products has originated different strategies such as "viral marketing", where
 inexpensive products as well as free services can be offered, generating 
streams of new consumers. In regards to Internet, this can be expressed in 
the form of new buttons providing either free mails or demo software, as 
dictated by Wilson marketing law: It is a question of giving and receiving \cite{1}. 
This way aggregate demand is sustained and therefore new Internet markets 
may survive, though market survival has been regarded as problematic ever
 since the last century and the absence of aggregate demand has led to Web 
Portals. Digital segregation inhibits access to Digital Economy, decreasing
 aggregate demand \cite{2} indicating future limitations for Internet expansion.
  Also aggregate demand  is very sensible to State
 intervention \cite{3} as well as to the induction of both "incorrect prices" and investors
 "incorrect expectations" in the banking system \cite{4}.      
   Social interaction is noticed through the influence received by individuals in the 
social networks, where an individual typically interacts with about eight to twelve
 members \cite{5}.
Notwithstanding there is a great variety of mass media, such as Internet and telephone, 
where high complexity communication is transmitted via face to face communication \cite{6}.

 Basu has developed algorithms to detect conversations among members of a group, in terms 
of synchronization, duration and frequency of the proximity between members of a group\cite{7}. 
This way, group connectors can be detected, namely those concentrating the transmission
 of information as well establishing some grounds to prove the influence of the
 environment on information diffusion. Nevertheless, there are fine effects affecting 
message transmission such as audience nodding frequency \cite{8}. 

   Using a simplified version of Asavathiratham model \cite{9} to reproduce first-neighbors 
interaction, 
Basu et. al. have developed a model that describes human interaction with the
 influence model, considering constant first-neighbors influence intensity as well as
 depending on first-neighbors state  \cite{10}..
   Campbell and Ormerod have studied the influence of social interaction on crime dynamics, 
with a model using a non linear system of differential equations. Due to the bifurcation of 
solutions emerging from the non linearity introduced by social interaction, time and location 
factors are more important than socio-economic status  \cite{11}. Employing a similar model, based 
on a system of differential equations, Campbell y Ormerod studied social interaction influence
 on English family structural evolution and found that the determinant factor for the English
 family structural evolution in the last three decades  was the increase of real salaries as well as
 the economic independence of female labor force. This behavior was explained as a result of 
non linearities introduced by social interaction in an equilibrium attractive solution \cite{12}.  

   Bonabeau et. al. have developed a first-neighbors interaction diffusion model, where 
 social hierarchy arises from population density fluctuations. Though initially every neighbor
 has the same probability to own the adjacent node, later a memory of the victories achieved is 
kept, periodically reducing up to a ten percent the value of this memory to reproduce the lack of social memory.
 This way the probability of winning a confrontation  is computed, taking on account the
 number of previous victories and the associated random origin of social hierarchy \cite{13}. 
   Stauffer et. al. studied the number of elements whose opinion remains unaltered, in 
Sznajd consensus model\cite{14}, using a multiple spin model where the sublattices are simultaneously
 actualized. They observed that the number of elements whose opinion remains unaltered decays 
on time following a power law $t^{- \theta}$, with an exponent
$\theta \sim \frac{3}{8}$  in the case of a linear chain and
 $\theta \sim \frac{1}{2}$ in the case of a square or 3D network
 \cite{15}. This is compared with the results obtained by Derrida et. al. for a squared network, with a value of 
$\theta \sim 0.2$ \cite{16}. 

   In this study, new product market penetration is analyzed, assuming that the market is already 
saturated by a given product, using an spintronic model that considers first-neighbors
 interaction. Social interaction and new product demand established by pre-campaign 
turn out to be the most important factors to ensure new product market penetration. This 
results in the creation of a new market, as new consumers are incorporated, originating an 
aggregate demand for a lapse of time.

\section{Methodology.}

   In this study, social environment  is simulated by one million
 individuals and first-neighbors interaction. Scale factor is taken
 on account considering that when new product production is increased,
 its production price is decreased. Besides, when the new product is used by 
more costumers, the price they are ready to pay is increased. Each element 
of the spherical network may assume three states : -1 when the element consumes
 the new product, 0 when the element does not consume the new product and
 1 when the element refuses the new product. Initially, every element is in state 0, 
namely not using the new product and is randomly distributed in a certain
 percentage.

This way system evolves randomly selecting an element, which considers the new
 product acceptation among its first-neighbors and this way takes a rational
 decision in terms of the maximum cost at which people is ready to pay for the new 
product and the price at which the new product is offered.
   Agents rationality and their market foresight are modeled following the model
 introduced by Hohnisch et. al. \cite{17}. A parabolic dependence is supposed between
 the minimum price at which the producer is ready to sell with and the number of produced units:
$p_{t}(x)= p_0 - qx + \alpha x^2$

   Besides, consumers are ready to pay a higher price when the new product has
 more consumers:
\begin{equation}
p_{a}(x)= p_0 + \mu x
\end{equation}

   Each potential consumer behaves as a rational agent, choosing the product
 with the best price and satisfying:
\begin{equation}
p_{a}(x)\ge p_t (x)
\end{equation}

This dynamics is followed to detect the determinant factor for new product penetration ,
 comparing the number of initial consumers and the acceptation coefficient $\mu$. Model
 constants were fixed elsewhere in a study on the adoption of a new corn variety at the state
 of Iowa \cite{18}. 
   When market penetration is simulated, new consumers clusters are built up. For infinite
 systems, there is a critical percolation probability, above which the probability of finding a
 percolation cluster is equal to 1 and for a value lower than this critical percolation the
 probability of finding a percolation cluster is 0. For finite networks, this transition is soft,
 i.e., the probability of finding a percolation cluster is not equal to 0 at any probability. 
Chayes et.al. have shown that the critical exponent for the scaling of finite clusters is equal
 to:$ \tau -2=1/2$ and a value of the critical exponent for the scaling of infinite clusters is 
equal to:  $\tau -2=1/3$ \cite{19}. In the percolation model de weight of a given configuration $C$ 
with n binding is:
$W(C)=p^n  (1-p)^{N-n}$
where N is the number of vertices in the network. Close to the percolation threshold, 
the critical behavior is characterized by the critical exponents:

\begin{equation}
P_{\infty} =1 - \sum  sn(s,p) \sim \envert{ (p - p_c) }^{\beta}
\end{equation}

and:
\begin{equation}
S(p)= \sum s^2 n(s,p) \sim \envert{p-p_c}^{\gamma}
\end{equation}

   The cluster distribution satisfies the following scaling relation:

\begin{equation}
n(s,p)=  s^ {-\tau}  f(p-p_c) s^{\sigma}
\end{equation}

therefore a power law is expected close to the critical point:

\begin{equation}
n(s,p)=  s^ {-\tau}  f(0) 
\end{equation}

   Using these relations, the following equation can be obtained:
\begin{equation}
\tau= 2+\frac {\beta}{\beta + \gamma} 
\end{equation}

and:
\begin{equation}
\sigma=\frac {1}{\beta + \gamma} 
\end{equation}

In 3D, the best estimation is $ \tau = 2.18$ and $ \sigma = 0.45$  \cite{20}. 
   In previous studies on percolation, critical exponents have been estimated for
 finite physical systems with values of the critical exponent in the range$ \tau \sim 2 - 3$\cite{21}.
 In the case of new product market penetration, a value of $ \tau = 2.88$   was obtained
 hereby, indicating the influence of social interaction.
  In the Fisher Liquid Droplet Model \cite{22}, the probability of obtaining a critical cluster,
 is given by:
\begin{equation}
n_A = q_0  A^{-\tau}
\end{equation}

with a proportionality constant $q_0$  that can be obtained using the first moment:
\begin{equation}
M_1 = \sum n_A
\end{equation}

of the normalized distribution, i.e.:

\begin{equation}
 M_1 = 1
\end{equation}

Hence, $q_0$ can be obtained from:
\begin{equation}
q_0 =\frac{1}{ \sum A^{1-\tau}}
\end{equation}

\section{Results}

   As shown in Fig. 1, as the number of new product consumers is increased, 
the maximum price that the consumers are ready to pay for the new product is
 also increased. And in a similar fashion, the price that the consumers are ready
 to pay for the old product is decreased. This happens for a given new costumers
 threshold generated from a pre-campaign where the new product is given as a
 present or  promoted by viral marketing techniques.
The results obtained indicate that the most important factor is the social interaction,
 as shown by the linearity of the Arrhenius curves with respect to the percentage of
 initial new consumers (Fig. 2). Therefore, new product market penetration depends
 on the capacity of previously introducing the new product in the population. This
 can be achieved using techniques as those used by viral marketing, where a
 new product acceptation threshold is reached, independently of its technical characteristics.
   Arrhenius plot has been fitted in terms of new consumers cluster size, obtaining 
a critical exponent equal to 2.88 (Fig. 3). Since Arrhenius plot is fitted selecting intermediate
 size clusters, the phase transition is related to the onset of new consumers clusters
 with an intermediate size, which is known to be of approximately 8 to 12 consumers.

\section{Conclusions}

   The simulation suggests that a pre-campaign previous to the introduction
 of the new product is necessary to ensure new product market penetration, 
independently of its technical characteristics. The importance of social
 interaction is expressed in terms of the need to reach an adequate threshold of new
 consumers to displace the older product. This indicates the
 convenience of using viral marketing techniques with the purpose of attaining
 this consumption threshold before the new product distribution.
 This study on the built up of new consumer clusters indicates a phase transition
 with a critical exponent equal to 2.88, which is different of those belonging to
 other physical systems.
   Author acknowledges partial support from UAM-Azcapotzalco and the
 access to computational facilities of Intensive Computation Lab of UAM-A.

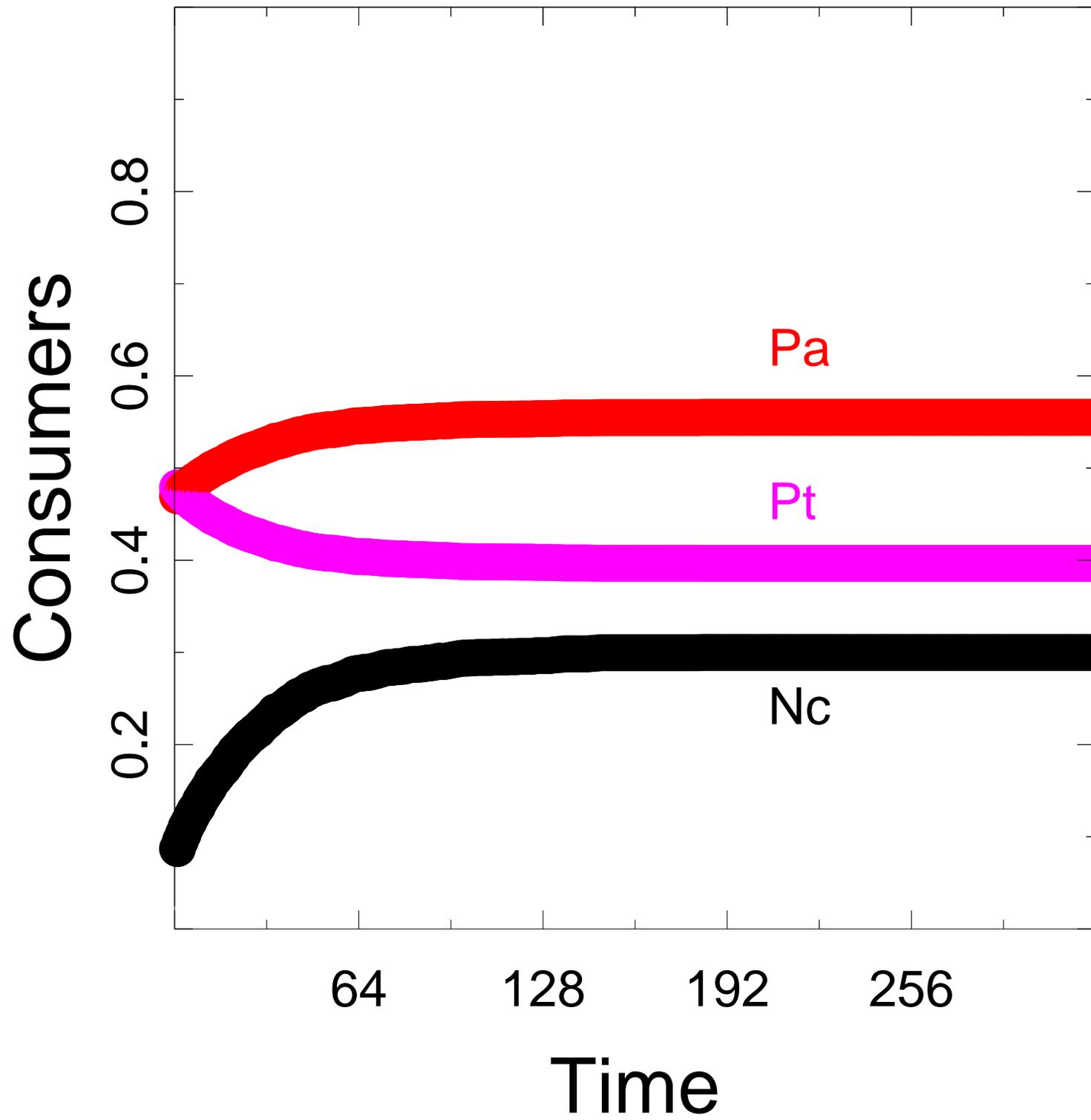
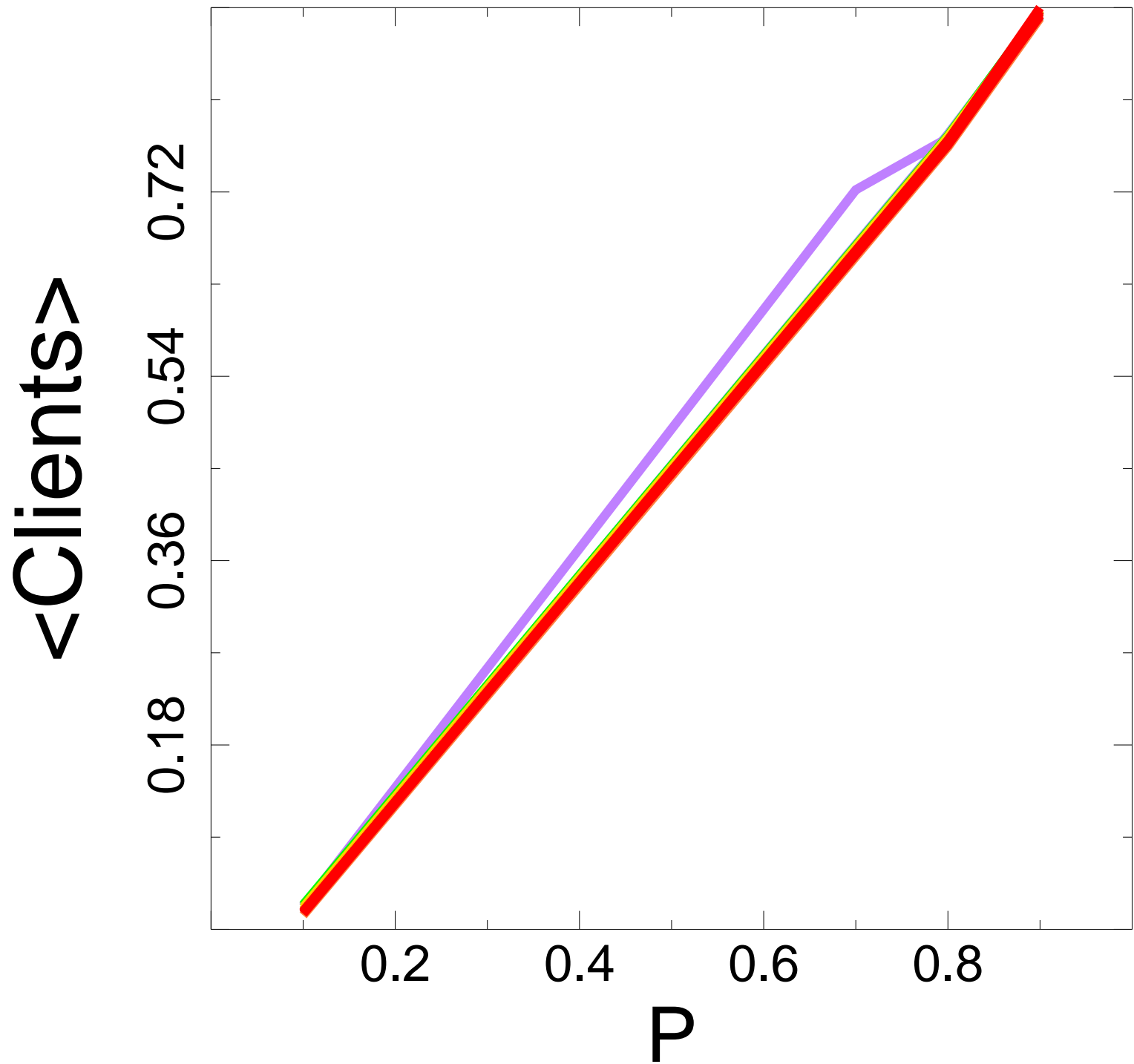
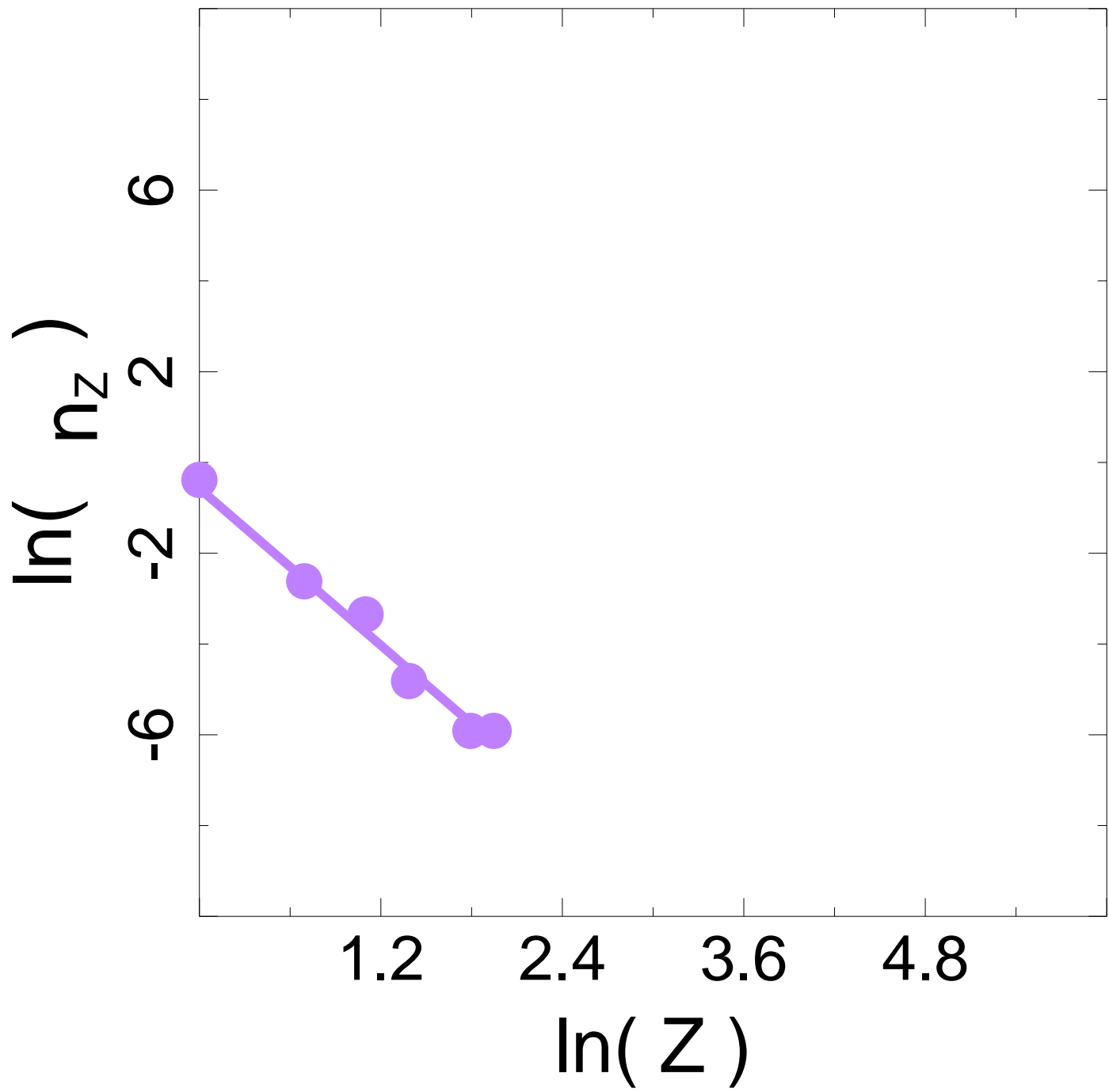
\end{document}